\documentclass[aps,prb,twocolumn,superscriptaddress,showpacs]{revtex4}
\usepackage{graphicx}
\usepackage{dcolumn}
\usepackage{bm}

\newcommand{\bperp}{B$_{\perp}$}
\newcommand{\bone}{B$_1$~}
\newcommand{\bth}{B$_{th}$~}
\newcommand{\bre}{B$_{re}$~}
\newcommand{\pfsix}{(TMTSF)$_2$PF$_6$}

\newcommand{\tone}{{1/T$_1$~}}

\newcommand{\tmt}{(TMTSF)$_2$ClO$_4$}

\begin{document}

\preprint{APS/123-QED}

\title{$^{77}$Se NMR investigation of the field-induced spin-density-wave transitions in (TMTSF)$_2$ClO$_4$}

\author{L. L. ~Lumata}
\affiliation{Department of Physics  and National High Magnetic
Field Laboratory, Florida State University, Tallahassee, FL 32310
USA}
\author{J. S. ~Brooks}
\affiliation{Department of Physics and National High Magnetic
Field Laboratory, Florida State University, Tallahassee, FL 32310
USA}
\author{P. L. ~Kuhns}
\affiliation{Department of Physics  and National High Magnetic
Field Laboratory, Florida State University, Tallahassee, FL 32310
USA}
\author{A. P. ~Reyes}
\affiliation{Department of Physics  and National High Magnetic
Field Laboratory, Florida State University, Tallahassee, FL 32310
USA}

\author{S. E. ~Brown}
\affiliation{Department of Physics and Astronomy, University of
California at Los Angeles, Los Angeles, CA 90095 USA}

\author{H. B. ~Cui}
\affiliation{Department of Physics  and National High Magnetic
Field Laboratory, Florida State University, Tallahassee, FL 32310
USA}

\author{R. C. ~Haddon}
\affiliation{Department of Chemical and Environmental Engineering,
University of California, Riverside, CA 92521 USA}
\date{\today}

\begin{abstract}
Complementary $^{77}$Se nuclear magnetic resonance (NMR) and
electrical transport have been used to correlate the spin density
dynamics with the subphases of the field-induced spin density wave
(FISDW) ground state in \tmt. We find that the peaks in the
spin-lattice relaxation rate 1/T$_1$ appear within the metal-FISDW
phase boundary and/or at first-order subphase transitions. In the
quantum limit above 25 T, the NMR data gives an insight into the
FISDW electronic structure.
\end{abstract}


\maketitle

The effects of high magnetic fields on the quasi-one- and
two-dimensional electronic structure of organic conductors is a
rich area of investigation.\cite{IshiguroYamaji} In Bechgaard and
related salts\cite{Oshima,Biskup} which remain metallic at low
temperatures, a magnetic field applied parallel to the least
conducting direction (perpendicular to the conducting chains)
produces a field-induced spin density wave (FISDW) ground
state.\cite{Chaikin1}  A simple description of this effect is that
the magnetic field decreases the amplitude of the lateral motion
of the carriers as they move along the conducting chains, thereby
making the electronic structure increasingly more one-dimensional.
Hence eventually 1D instabilities become favorable. In reference
to Fig. 1, a nested quasi-1D Fermi surface is induced at a
second-order phase boundary where a FISDW gap opens above a
threshold field B$_{th}$. \cite{Pesty} Due to quantization of the
nesting vector Q=(2k$_F$$\pm$N2$\pi$/$\lambda$,$\pi$/b) (where
$\lambda$=h/ebB), increasing magnetic field produces a first-order
``cascade" of FISDW subphases. In the quantum limit, the optimum
nesting vector (where N=0) yields the final FISDW state in the
case of (TMTSF)$_2$PF$_6$ which has a single quasi-one-dimensional
Fermi surface (Q1D FS).\cite{Kang}

However, (TMTSF)$_2$ClO$_4$ experiments show additional phase
boundaries in the quantum limit,
\cite{osada1,Naughton1,McKernan,UjiRO,Chung}.  Since the ordering
of the tetrahedral ClO$_4$ anions below 24 K doubles the unit cell
along the inter-chain direction b$^{\prime}$, zone folding
produces two Q1D FS sheets, leading to complex high field
behavior.\cite{UjiRO,Chung} Recently a model\cite{McKernan2} has
been proposed where both FS sheets are gapped at the Fermi level
E$_F$ in the FISDW region, but above the ``re-entrant" phase
boundary B$_{re}$,\cite{Naughton1} only one of the sheets is
gapped at E$_F$. This leads to an explanation for the oscillatory
sign reversal of the Hall effect \cite{UjiHall} above B$_{re}$.

\begin{figure}[tbp]
\begin{center}
\includegraphics[width=3.0in]{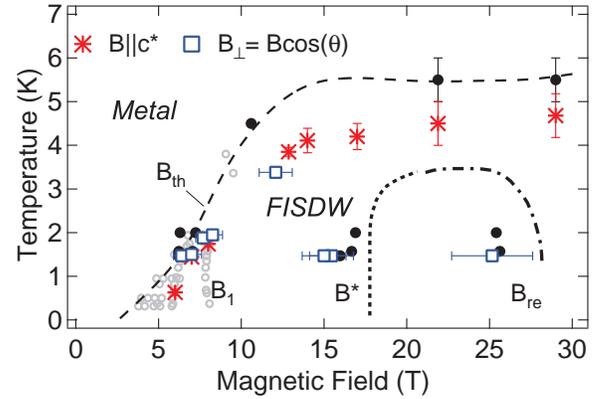}
\caption{(Color online). Phase diagram of (TMTSF)$_2$ClO$_4$ for
B$\parallel$c* derived from previous
reports\cite{McKernan,Chung,Naughton,Pesty} (dashed lines)
including a summary of the observed $^{77}$1/T$_1$ peaks
(asterisks and open squares) and the corresponding features in the
transport measurements (dark and gray circles) from this work.
(Field labels defined in text.)} \label{fig:nmr1}
\end{center}
\end{figure}

In this report we present complementary measurements of electrical
transport and pulsed $^{77}$Se ($\gamma$=8.13 MHz/T) NMR in \tmt,
both for variable frequency and magnetic field, and by rotating
the sample in the b$^{\prime}$-c* plane at constant frequency and
magnetic field (see inset of Fig. 3(c) for definitions of crystal
axes and field direction $\theta$). The orbital nature of the
FISDW transitions \cite{Boebinger} allows a continuous sweep of
the perpendicular field \bperp=B$\cos$$(\theta)$ from 0 T for
B$\parallel$b$^{\prime}$ to 30 T (in the present case) for
B$\parallel$c*, thereby accessing the entire metallic and FISDW
field range by sample rotation at constant NMR frequency.

\begin{figure}[tbp]
\linespread{1}
\par
\includegraphics[width=3.3in]{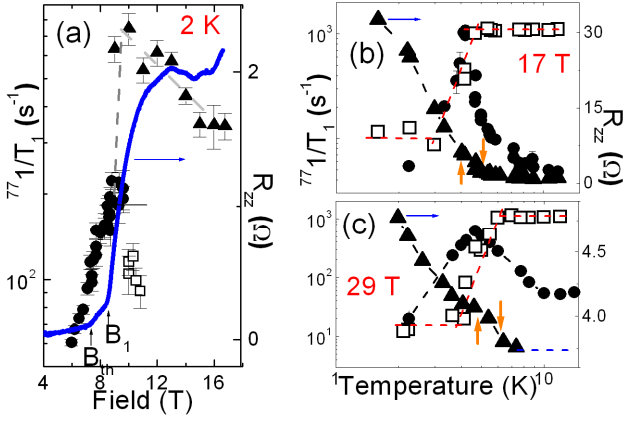}
\par
\caption{(Color online). a) Field-dependent 1/T$_1$ and resistance
at 2 K ($\theta = 25^o$). Solid circles and open squares: metallic
pulses; solid triangles: enhanced FISDW pulses.(b) and (c)
Temperature dependence of 1/T$_1$ (solid circles), R$_{zz}$ (solid
triangles), and the integrated NMR intensity with Boltzmann factor
correction (open squares) for B$\parallel$c*. The down arrow
indicates the upturn in R$_{zz}$ and the up arrow indicates the
peak 1/T$_{1}$.} \label{fig:2}
\end{figure}

A single crystal of (TMTSF)$_2$ClO$_4$ was inserted into a
miniature NMR coil with the a-axis aligned parallel with the coil
axis. Gold wires (12$\mu$m) were attached with carbon paint for
four-terminal ac resistance measurements with a current of 1
$\mu$A applied along the c*-axis. The sample and coil were mounted
on a probe with single-axis goniometer, and cooled at a rate of 30
mK/min from 30 K to 18 K to allow anion ordering. For variable
field measurements, the NMR frequency was changed for each field,
and for rotation measurements at fixed field B$_{max}$ in the
b$^{\prime}$-c* plane, the frequency was fixed at
$\gamma$B$_{max}$.


The spin density wave nature of the ground state has been
described in the Bechgaard salts by NMR
studies.\cite{Takahashi,Delrieu} The characteristic NMR lineshape
in the SDW phase involves multiple peaks due to the
antiferromagnetic nature of the local fields, and the peak
separations  vary systematically with field orientation due to
changes in the dipolar coupling with respect to the
(TMTSF)$_2$ClO$_4$ donor axes.\cite{Takigawa1986, Zhang2005} 


NMR pulse optimization involved $\pi/2$-$\pi/2$ pulse trains to
obtain the maximum NMR intensity at the optimum pulse widths
$\tau_{M}$ in the metallic and $\tau_{sdw}$ in FISDW phases.
Typically, in the metallic state the optimum $\pi/2$ pulse width
was $\tau_{M}$ = 1 $\mu$s while in the FISDW region $\tau_{sdw}$
varied from 50 ns to 500 ns. Since the same pulse power level (12
W) was used for all measurements, the rf enhancement
factor\cite{Takigawa} was obtained from the relation
$\eta\equiv\tau_{M}/\tau_{sdw}$. The spin-lattice relaxation rate
1/T$_1$ was defined using a single-exponential form for the
magnetization recovery, and varied by less than 10$\%$ over the
spectrum in all cases. In the FISDW, there is a slight deviation
(approx 5\% in the initial slope) from the exponential recovery,
not uncommon for these systems.\cite{Zhang2005} Long recycle times
(0.5 to 5 s) were used to avoid sample heating.


\begin{figure}[tbp]
\linespread{1}
\par
\includegraphics[width=2.9in]{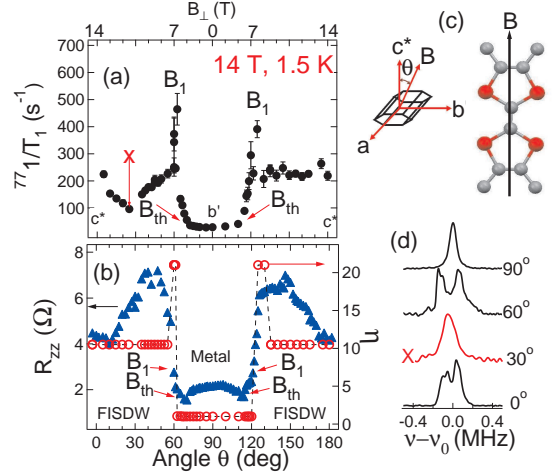}
\par
\caption{(Color online). Angular dependent NMR and electrical
transport at 14 T and 1.5 K in \tmt. (a) 1/T$_1$ vs. field
orientation. (b) Corresponding magnetoresistance and enhancement
$\eta$ at 14 T. (c) Schematic of the crystal axes and orientation
of the TMTSF molecule at point X where the dip in 1/T$_1$ occurs.
(d) Representative spectra at various angles (X represents the
spectrum at the dip in 1/T$_1$).}\label{fig:nmr3}
\end{figure}

We first discuss  results  at constant angle where the field (and
frequency) were changed to access the different FISDW phases. In
Fig. \ref{fig:2}(a) the magnetic field dependent 1/T$_1$ and the
corresponding c*-axis resistance (R$_{zz}$) are shown for 2 K. For
constant temperature, as the metal-FISDW transition approached and
\bth is crossed, 1/T$_1$ gradually increases. It is not until the
predominant first-order sub-phase transition B$_1$ is reached that
1/T$_1$ exhibits a maximum. For further increases in field
1/T$_{1}$ decreases. The metallic pulses were optimum when
$B<B_1$, but at higher fields the shorter $\tau_{sdw}$ pulses were
necessary to follow the signal into the FISDW phase. In Fig.
\ref{fig:2}(b) and \ref{fig:2}(c), the variation of 1/T$_1$ with
temperature is shown, along with R$_{zz}$ and the Boltzmann factor
normalized NMR intensity. The temperature dependence of 1/T$_1$
generally follows a critical fluctuation behavior \tone $\approx
(T-T_{SDW})^{-1/2}$ above the peak , and a power law behavior
(\tone $\approx T^{1.2}$) at lower temperatures. The spin density
is constant in the metallic state, and drops exponentially
starting at the second-order phase boundary. Simultaneous
resistance measurements show the onset of semimetallic behavior at
this same temperature. We emphasize here that the peaks in 1/T$_1$
occur at temperatures that are systematically below the
second-order phase boundary, as indicated by the resistance
anomaly and the signal intensity. The positions of the peaks in
1/T$_{1}$, and the corresponding phase boundaries from transport
measurements, are presented in Fig. 1 for all field-dependent
data.



The angular-dependent data shown in Fig. 3 at 14 T (118.32 MHz)
and 1.5 K provide more detail about the second-order B$_{th}$ and
first-order B$_{1}$ FISDW transitions. In the angle range
$70^{0}<\theta<110^{0}$, the sample is metallic, which is also
shown in magnetoresistance (MR) and $\eta =1$. There is a slight
increase (10 s$^{-1}$) in 1/T$_{1}$ as the field rotates away from
the b$^{\prime}$-axis in the metallic state. When B$_{\perp}$
reaches B$_{th} = 6.25 T$, there is an increase in 1/T$_{1}$
corresponding to a sharp feature in the resistance. Note however
that for B$_{th} < B_{\perp} < B_{1}$ , $\eta$=1 as in the
metallic phase. As in Fig. 2a, the peak in 1/T$_{1}$ occurs at
B$_{1}$ where $\eta$$\approx$20. Deep in the FISDW phase, the NMR
lineshapes broadened and have a double-peaked structure. Since the
a-axis used for the rotation data is also the symmetry axis for
the hyperfine coupling, anisotropy in 1/T$_{1}$ is indiscernible
in the metallic phase,\cite{Wu} and negligible in the FISDW phase
(a few percent) compared with the large variation in 1/T$_{1}$ at
the different phase boundaries. In the angular dependence there is
a special symmetry position at $\theta$$\approx$25$^{0}$ where the
magnetic field is parallel to the x-axis of the TMTSF donor
molecule. Here 1/T$_{1}$ exhibits a dip (marked X in Fig. 3) and
the lineshape narrows to a single peak. This dip feature has been
previously reported for \pfsix,\cite{Takahashi} and is not related
to the magic angle (MA) effects \cite{Lebed} reported in
(TMTSF)$_2$ClO$_4$.\cite{Naughton}. Although FISDW data is always
above the spin-flop field ($<$0.5 T), there is a spin rotation
from the a-axis to the c*-axis when the field direction approaches
B$\parallel$c*, which has been associated with the dip
phenomena.\cite{Takahashi} Likewise, there is no evidence for the
MA effects in the NMR signal, in accord with previous studies on
\pfsix.\cite{Wu}

\begin{figure}[tbp]
\linespread{1}
\par
\includegraphics[width=3.2in]{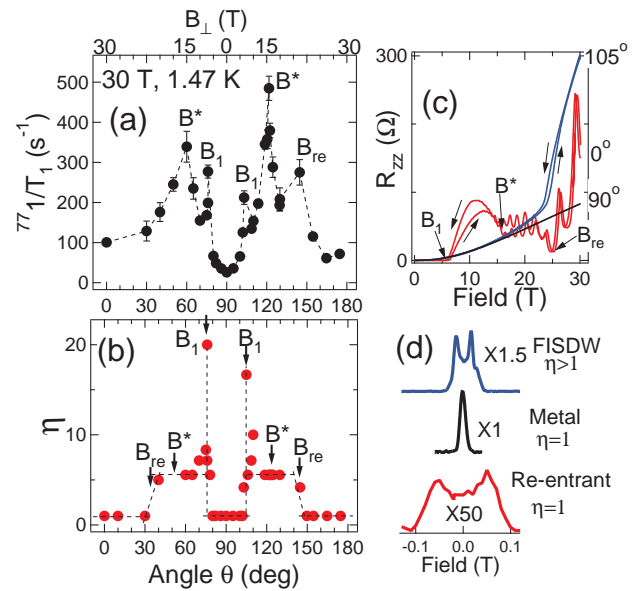}
\par
\caption{(Color online). Angular dependent NMR and electrical
transport at $B=30T$, 1.47K. (a) Metallic and FISDW phase
transitions revealed in angular-dependent 1/T$_1$. (b) The
enhancement vs. $\theta$. Of special note is that $\eta = 1$ above
B$_{re}$. (c) Representative MR at different angles at 1.47 K. For
each trace, the NMR measurement was made at 30 T , and hence above
\bre for $\theta = 0^o$, above B$_1$ for $\theta = 105^o$, and in
the metallic phase for $\theta = 90^o$. (d) Corresponding
fieldswept spectra at $\nu_{0}=243.9$ MHz (30 T) at different
phases.} \label{Fig4}
\end{figure}

In  Fig. \ref{Fig4} the high field (B$_{max}$= 30 T)
angular-dependent NMR results for 1/T$_{1}$, B$_{1}$, B*,
B$_{re}$, and $\eta$ are shown along with the corresponding MR
data at T=1.47 K and 243.9 MHz. As in Fig. \ref{fig:nmr3},
rotation away from the metallic phase at B$\parallel$b$^{\prime}$
(i.e. for increasing B$_{\perp}$) causes \tone to increase as \bth
is entered, but it is not until the first-order boundary \bone is
crossed that 1/T$_{1}$ reaches a maximum. Since B$_{max}$=30 T,
the full field range in B$_{\perp}$ can be accessed, and
additional peaks in \tone are observed for \bperp=B* in the range
15 to 17 T (which corresponds to a characteristic feature in the
MR seen in many experiments \cite{Chung}), and also for
\bperp=B$_{re}$. Of specific note is the corresponding behavior of
$\eta$ which falls from 5 to 1 when the \bre boundary is crossed
for both positive and negative field directions. The main changes
in spectral linewidth due to the internal field occur at the
metal-FISDW transition, and not at the subphase transitions. Hence
the enhancement factor and increase in 1/T$_1$, and not a
significant change in the internal field, characterizes the
B$_{re}$ phase boundary. For $\theta\approx 25^{o}$ in Fig. 4(a)
the feature corresponding to \bperp=\bre is obscured, most likely
because it occurs where the dip (see Fig. 2) in \tone appears. The
results from Fig. 4 are summarized in Fig. 1.

Most significant in the present work is the location of the peaks
in \tone in the different FISDW phases, and the behavior of the
NMR signal above the re-entrant phase boundary.

First, for \tone data at constant field, the peaks in \tone appear
at temperatures as much as $30\%$ lower than the second order
phase boundary (Fig.1).  Hanson \emph{et al} have previously
suggested\cite{Hanson} that the true signature of the second order
onset of the FISDW occurs at the \tone peak, and not at the higher
temperature onset of transport anomalies. However, the rapid
oscillation (RO) behavior (see, e.g., Fig.4(c)) is very sensitive
to the Fermi surface nesting conditions, and a transition between
Stark interference behavior (metallic) and anomalous RO behavior
(FISDW) is coincident with the second order phase
boundary.\cite{UjiStark,UjiRO,Chung} The spin density, and the
onset of semimetallic behavior in R$_{zz}$ also begins to change
at this boundary (Fig. 2c); see also Fig.6(a) in Ref.[9]). We
therefore assert that the FISDW phase boundary, i.e. the second
order line, appears when the FS nesting (and gap) first appears,
that is, \emph{above} the temperature where the peaks in \tone
appear. From Fig.2(b,c), this effect occurs approximately at the
inflection point of the drop in the spin density ($50\%$). In
contrast, we note that in (TMTSF)$_2$PF$_6$ where under ambient
conditions a SDW transition occurs at 12 K, the resistance shows a
sharp increase \cite{Bechgaard} which is nearly coincident with a
peak in 1/T$_{1}$,\cite{Valfells} with an uncertainty of less than
$10\%$. Hence there appears to be a difference in the way the
order parameter develops below the SDW and the FISDW phase
boundaries. Critical slowing of the fluctuations near a SDW
transition ordinarily leads to a peak in relaxation rate at $T_c$.
Below $T_c$, relaxational (dynamical) effects have been
observed\cite{Clark}, which cannot be ruled out. However, a
density of states effect similar to the Hebel-Slichter peak seen
in superconductors\cite{Charlie} is also possible.

Second, in both field dependent and rotation experiments, we find
that \tone exhibits peaks within the second-order phase boundary,
in particular at the first-order phase lines B$_1$, B*, and
B$_{re}$. Since the first order transitions involve transitions
between the different sub-phases, domain wall effects may cause
the peaks in \tone as these boundaries are crossed.\cite{Brownpf6}

Third, in the rotation experiments the enhancement parameter
$\eta$ was monitored in detail. The effect is associated with
rf-induced displacement or depinning of the SDW phase by the
electric fields. There is no enhancement ($\eta$=1) in the
metallic phase, and in the FISDW phase $\eta$ will increase
according to the ability of the electric field to modulate the
condensate.\cite{Brown2} The electric fields associated with
de-pinning are typically 5 mV/cm or less.\cite{Wong1, Osada1987}
In the present case, we estimate the ac electric field in the NMR
coil to be of order 1 to 10 V/cm (see also previous
estimates\cite{Takigawa}), well above the de-pinning field.
Notably, at the transition \bre, the enhancement factor drops to
unity, but the NMR spectrum is still double-peaked and
characteristic of antiferromagnetic structure (Fig.4(d)). Based on
the recent band model\cite{McKernan2} in the re-entrant phase, the
FISDW will be associated with one of the bands (FS1 - giving rise
to the SDW spectrum), but the other band (FS2) is metallic at the
Fermi level (reducing the effective electric field). Hence we
propose that the electric field in the half metallic phase above
\bre is attenuated, thereby reducing $\eta$.

In summary, we have correlated the transport features which
describe the FISDW phase diagram of (TMTSF)$_{2}$ClO$_{4}$ with
$^{77}$Se NMR. We find that the peaks in \tone occur within (not
at) the second order phase boundary. Furthermore angular dependent
measurements facilitate the crossing of FISDW subphase boundaries
at constant temperature where peaks in \tone are also observed,
indicating changes in the nesting configurations at these
transitions. At high fields, the drop in the rf enhancement
parameter upon crossing the re-entrant phase boundary is
consistent with the expectation that only one FS sheet is nested
above B$_{re}$. We expect that this electronic configuration (only
one FS sheet may be nested) is also the case in the broader region
between the second order phase boundary and the underlying first
order phases at lower temperatures.

We thank V.F. Mitrovic for helpful suggestions. This work was
supported in part by NSF DMR-0602859 (JSB) and DMR-0520552 (SEB),
and performed at the National High Magnetic Field Laboratory,
supported by NSF DMR-0084173, by the State of Florida, and the
DOE.

\end{document}